\documentclass[twocolumn,preprintnumbers,amsmath,amssymb]{revtex4}
\usepackage{graphicx}
\usepackage{dcolumn}
\usepackage{amsmath}
\usepackage{bm}
\DeclareGraphicsExtensions{.pdf,.png,.jpg}
\usepackage{epstopdf}
 
\begin{document}
 
\title{Electronic origin of structural transition in 122 Fe based superconductors}
\author{Haranath Ghosh, Smritijit Sen and Abyay Ghosh}
\affiliation{Human Resources Development Section, Raja Ramanna Centre for Advanced Technology, 
Indore -452013, India. \\ and Homi Bhabha National Institute, Anushakti Nagar, Mumbai 400094, India.}
\date{\today}
\begin{abstract}
Direct quantitative correlations between the orbital order and orthorhombicity is achieved in a number of Fe-based superconductors of 122 family. The former (orbital order) is calculated from first principles simulations using experimentally determined doping and temperature dependent structural parameters while the latter (the orthorhombicity) is taken from already established experimental studies; when normalized, both the above quantities quantitatively corresponds to each other in terms of their doping as well as temperature variations. This proves that the structural transition in Fe-based materials is electronic in nature due to orbital ordering.  An universal correlations among various structural parameters and electronic structure are also obtained. Most remarkable among them is the mapping of two Fe\--Fe  distances in the low temperature orthorhombic phase, with the band energies E$_{d_{xz}}$, E$_{d_{yz}}$ of Fe at the high symmetry points of the Brillouin zone. The fractional co-ordinate $z_{As}$ of $As$ which essentially determines anion height is inversely (directly) proportional to Fe-As bond distances (with exceptions of K doped BaFe$_2$As$_2$) for hole (electron) doped materials as a function of doping. On the other hand, Fe-As bond-distance is found to be inversely (directly) proportional to the density of states at the Fermi level for hole (electron) doped systems. Implications of these results to current issues of Fe based superconductivity are discussed. 
\vspace{1pc}
\end{abstract}
\maketitle

\section{Introduction}
The discovery of the iron-based superconductors modified the notion of uniqueness of  high temperature 
superconductivity in cuprates and is interesting even after eight years of its discovery 
not only because they have the ability to exhibit superconductivity at very high transition temperatures, 
but also because it provides a rich prototype interplay of various degrees freedom ; the lattice 
and the electronic charge, spin, and orbital degrees of freedom all have intriguing roles. The 122 
family of Fe based superconductors are metals but with multi-band and multi-orbitals at all doping ranges in their
respective phase diagram, this should be contrasted with the case of  high temperature copper-oxide superconductors
 which are Mott-insulators when undoped. 
The undoped and doped Fe-based superconductors undergo a structural transition from a high-temperature tetragonal phase to a low-temperature orthorhombic phase which may coincide with the magnetic transition or the structural 
transition precedes the magnetic transition. These compounds below the magnetic transition are believed to be antiferromagnetic like spin density wave (SDW) metal. 
These transitions are suppressed with electron or hole doping (or pressure ) 
and superconductivity (SC) eventually appears beyond a certain doping concentration with high critical 
transition temperature along with coexistence between both SDW and SC in certain cases. 
Presumably, this lead to a number of initial investigations  mainly focused on the interplay and competition between a stripe-like antiferromagnetism (where the magnetic ordering wave-vector 
is $(\pi, 0)$ or $(0, \pi)$ in the 1Fe/cell notation) and superconductivity. 
It is popularly believed that an anti-s wave (s$\pm$) nodal superconducting pairing symmetry \cite{Kazuhiko,Mazin,Hirschfeld} is mediated through the fluctuations associated with the stripe SDW state. This cannot however be valid with certainties to many of the Fe- based superconducting families \cite{Reid}.

A number of studies on ``nematicity'' or strongly anisotropic in-plane transport properties like resistivity etc. which is also viewed as ``preferential transport'' in the 122 systems \cite{Chu} in the orthorhombic phase or in tetragonal phase in presence strain field, created a lot of attention leading to a new phase called `nematic phase'  in the phase diagram of these materials. A nematic phase of matter is one in which the order parameter for a transition breaks rotational symmetry  but time-reversal symmetry is preserved similar to the order parameter in the nematic phase of liquid crystals. By analogy, the orthorhombic state in FeSCs has been called a ‘nematic state’. Such phases were well studied in classical soft matter systems \cite{Prost}, but not in their quantum counterparts in electronic systems. However, strongly correlated electron systems such as quantum Hall systems, cuprates, bilayer ruthenates \cite{Fradkin} widely postulated their existence. In a crystalline lattice,
a nematic phase corresponds to the breaking of discrete rotational symmetry, and in the context of the Fe based SC, 
this order parameter is non-zero in the orthorhombic phase where the C$_4$ symmetry 
is broken at the structural transition, T$_S$. As mentioned earlier, in certain systems, the structural and the magnetic  transition are identical (T$_S$=T$_N$), and since the magnetic order (stripe SDW) by itself breaks 
C$_4$ symmetry, the primary order parameter for nematicity appears to be magnetic one.

In the nematic phase actually all the three parameters (a) structural distortion (phonon driven), 
(b) charge/orbital order (n$_{d_{xz}}$, n$_{d_{yz}}$ being different),  (c) spin order (static spin susceptibility 
along q$_x$ and q$_y$ being different) are non-zero \cite{Song}, no mater what drives the nematic instability. 
A bilinear combinations of these order parameters (a, b, c) that break tetragonal symmetry are 
invariant under symmetry transformations and forms essential part of Landau Free energy. 
As a result, from principle of minimization one order parameter induces 
the other. And that is precisely the experimental challenge as how to determine the primary 
order parameter responsible for nematic transition. Consequently, the issue of nematicity 
is more crucially posed for those systems where the structural transition precedes the magnetic one (T$_S\>$T$_N$), 
leaving a finite temperature interval where C$_4$ symmetry is broken but the material remains paramagnetic 
\cite{Ni,Chu2,Luetkens,Parker}. The most notable example being FeSe where only a structural transition \cite{Margadonna,McQueen} is detected (the difference between the T$_s$ and T$_N$ being the largest) and the system remains paramagnetic till its SC phase \cite{Hsu,Imai}, indicating that nematic degrees of freedom are not necessarily magnetic ones but probably orbital one.
The microscopic origin of the nematic order is debatable with a few competitive probable scenarios \cite{FernandesN}. 
One scenario is that the structural transition is an anharmonic lattice potential driven instability, 
so the lattice orthorhombicity would be the primary order parameter.
A second scenario is that the C$_4$ symmetry  is  broken primarily by electronic interactions so that lattice
degrees of freedom are secondary order parameters that passively follow the symmetry breaking induced by the electronic interactions hence the primary order parameter is electronic in origin. One such scenario is the 
spin-nematic transition whereby the spins of the two Fe sublattices phase-lock, this breaks C$_4$ 
symmetry, without developing any spontaneous magnetization, \textit{i.e.}, without breaking time reversal 
symmetry \cite{WZhang,Chandra,Fang,Xu,Qi,Fernandes,Paul,Cano,Fernandes2,Fanfarillo,Sen}. Another 
possible candidate in this scenario is ferro-orbital ordering \cite{Kruger,Lee,Chen,Lv,Onari,Sen}, 
where below nematic transition either the occupations or the hopping matrix elements (or both) of 
the d$_{xz}$ and the d$_{yz}$ orbitals of Fe become inequivalent. Apart from the above 
two electronic scenarios, other possibilities include a d-wave Pomeranchuk instability \cite{Zhai}, 
in which the Fermi surfaces undergo symmetry-breaking distortions due to interaction effects.

From above discussions it turns out that the primary step would be to show beyond doubt as to 
whether the structural transition is electronic in origin or lattice driven. This is the main purpose of 
the present work. In this work we consider several Fe-based superconducting materials from 122 
family to show a direct quantitative correlations between the orbital order and orthorhombicity. 
The orbital order is calculated from electronic band structures through first principles 
simulations using doping and temperature dependent experimentally determined structural parameters. The orthorhombicity parameter is taken from already established experimental studies (refs \cite{Ni,Acta,Allred,Avci2}) as a function or various doping concentrations. When such variations are normalized, be it thermal variation 
or doping dependent, both the above quantities (orbital order and orthorhombicity) 
quantitatively corresponds to each other in terms of their doping as well as temperature 
variations. This is shown universally for hole doped, electron doped and iso-electronic
 substituted materials. The orbital order being (the difference in band energies E$_{d_{xz}}$, E$_{d_{yz}}$ 
 at high symmetry k-points) derived from electronic band structure, 
 its complete matching with orthorhombicity proves that the structural transition 
 in Fe-based materials is electronic in nature. This is thus a direct proof that 
 orthorhombic distortion is {\it not} due to lattice but an electronic one. A first step to reject one 
 of the order parameters that may lead to nematicity. That leaves orbital fluctuation or spin order 
 as primary order for nematic phase. 
 An universal correlations among various structural 
 parameters and electronic structure are also obtained. Most remarkable among them is the 
 mapping of two Fe-Fe distances in the low temperature orthorhombic phase, with the band energies 
 E$_{d_{xz}}$, E$_{d_{yz}}$ of Fe at the high symmetry points of the Brillouin zone. This correspondence 
 of band energies with two Fe-Fe distances is true for its variation 
 either with doping or temperature and is shown for a several undoped and doped 122 materials. Furthermore, the fractional co-ordinate z$_{As}$ of As which essentially determines the anion height is inversely (directly) proportional to Fe-As bond distances (with exception to K doped BaFe$_2$As$_2$) for hole (electron) doped materials as a function of doping. On the other hand, Fe-As bond-distance is found to be inversely (directly) 
 proportional to the density of states at the Fermi level for hole (electron) doped systems. 
 Chemical effect in terms of size of doping elements and nature of substitution, in-plane or 
 out of plane is employed to understand various doping dependent electronic structures. 
 Thus as a whole this work is a complimentary theoretical studies to the explanations of experimental findings \cite{Ni,Acta,Avci,Allred,Avci2}.
 Furthermore, indeed spin origin is also very important and we have seen various spin structures of 
 Fe clearly influence z$_As$ and the orbital order \cite{spinAIP}. Such calculation will be published shortly.
\section{Theory and Computational Method}
Our first principles electronic structure calculations are performed by implementing ultrasoft 
pseudopotential with plane wave basis set based on density functional theory \cite{CASTEP}. Electronic 
exchange correlation is treated within the generalised gradient approximation (GGA) using 
Perdew-Burke-Ernzerhof (PBE) functional \cite{PBE}. It should be mentioned here that the density funtional 
theory within local density approximation (LDA) as well as generalized gradient approximation (GGA) was
unable to optimize the experimental value of $z_{As}$ (fractional co-ordinate of As)
with desired accuracy \cite{Acta,sust,zAs,DJSingh,Zhang,Yin,Sn}. Experimentally, in all iron chalcogenides/pnictides (also in phosphides) the chalcogen/pnictogen 
tetrahedron is found very close to regular tetrahedron. First principles density functional theory 
using LDA and GGA result in z-direction compressed tetrahedron by 0.1 to 0.15\AA  ~difference. 
This is a structural mystery in the field of Fe-based superconducting materials that remains unsolved even eight years after its discovery. The origin of this discrepancy is ascribed to the presence of strong magnetic fluctuation, associated with Fe atoms
in these materials \cite{Mazin2}. Quantitatively, such underestimation of z$_{As}$ leads to mutual shifts 
in the Fe-3d bands upto 0.25 eV or slightly more which causes completely different behaviours in the structural parameters and huge difference in the FS from experimental situation \cite{pla}. In this respect, consideration of LDA and GGA exchange correlation functional do not make any difference.
To overcome such a situation, we resort to experimentally determined
lattice parameters {\it i.e.,} $a$, $b$, $c$ and $z_{As}$ as a funtion of doping as well as temperature,
instead of geometry optimized (total energy minimization) lattice parameters as inputs of our first principles 
electronic structure calculations. This is the best available solution at the moment because even finite temperature molecular dynamic simulation together with density functional study has failed this aspect and can not subtly obtain tetragonal to orthorhombic transition.


We use experimental orthorhombic (low temperature) as well as tetragonal (high temperature) lattice parameters 
$a$, $b$, $c$ and $z_{As}$ as input of our {\it ab-initio} electronic structure calculations. 
In order to dope the system theoretically, we use Virtual crystal approximation (VCA) \cite{Bellaiche,VCA} as well as
supercell approach (specially for Ru doping at Fe site).
Non-spin-polarized and spin polarised single point energy calculations 
are carried out for tetragonal phase with space group symmetry I4/mmm (No. 139) and 
orthorhombic phase with space group symmetry Fmmm (No. 69) respectively 
using ultrasoft pseudo-potentials and plane-wave basis 
set with energy cut off 600 eV and higher as well as
self-consistent field (SCF) tolerance as $10^{-7}$ eV/atom. 
Brillouin zone is sampled in the k space within Monkhorst–Pack scheme
and grid size for SCF calculation is chosen as per requirement of the calculation 
for different systems.
\section{Results and Discussions}
\par
\begin{figure}
  \centering
  \includegraphics [height=8cm,width=8.5cm]{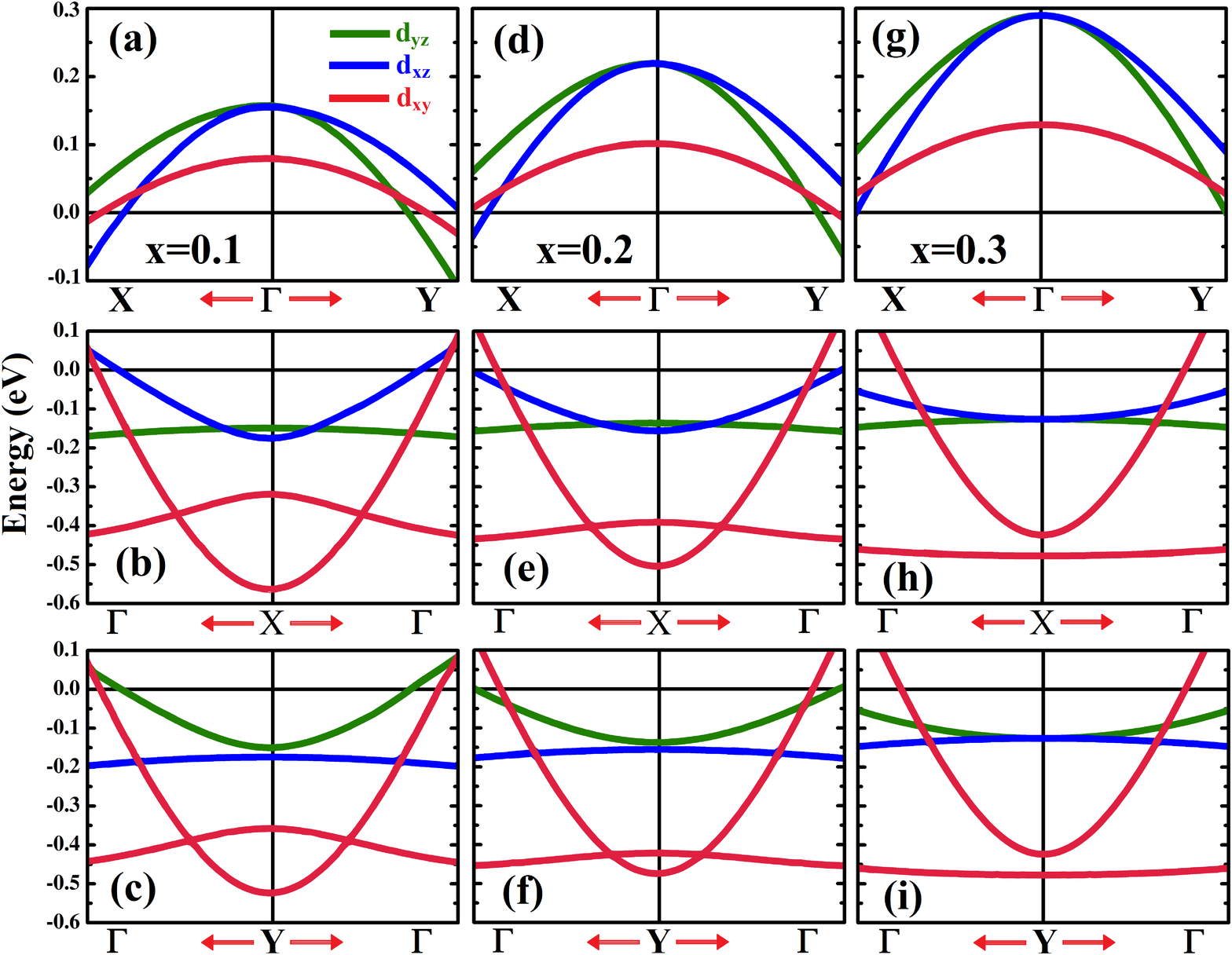}
  \caption{(Colour online) Calculated band structures of Ba$_{1-x}$K$_x$Fe$_2$As$_2$ system within VCA method 
    in the low temperature orthorhombic phase 
    around $\Gamma$ (1st row), X (2nd row) and Y (3rd row) points for 
    (a, b, c) $x=0.1$ (d, e, f) $x=0.2$ and (g, h, i) $x=0.3$ respectively. Fermi level is denoted by 
    horizontal black line at zero energy. Orbital orders around all these high symmetry points are calculated.}
  \label{KBS}
  \end{figure}

\begin{figure}
  \centering
  \includegraphics [height=8cm,width=8.5cm]{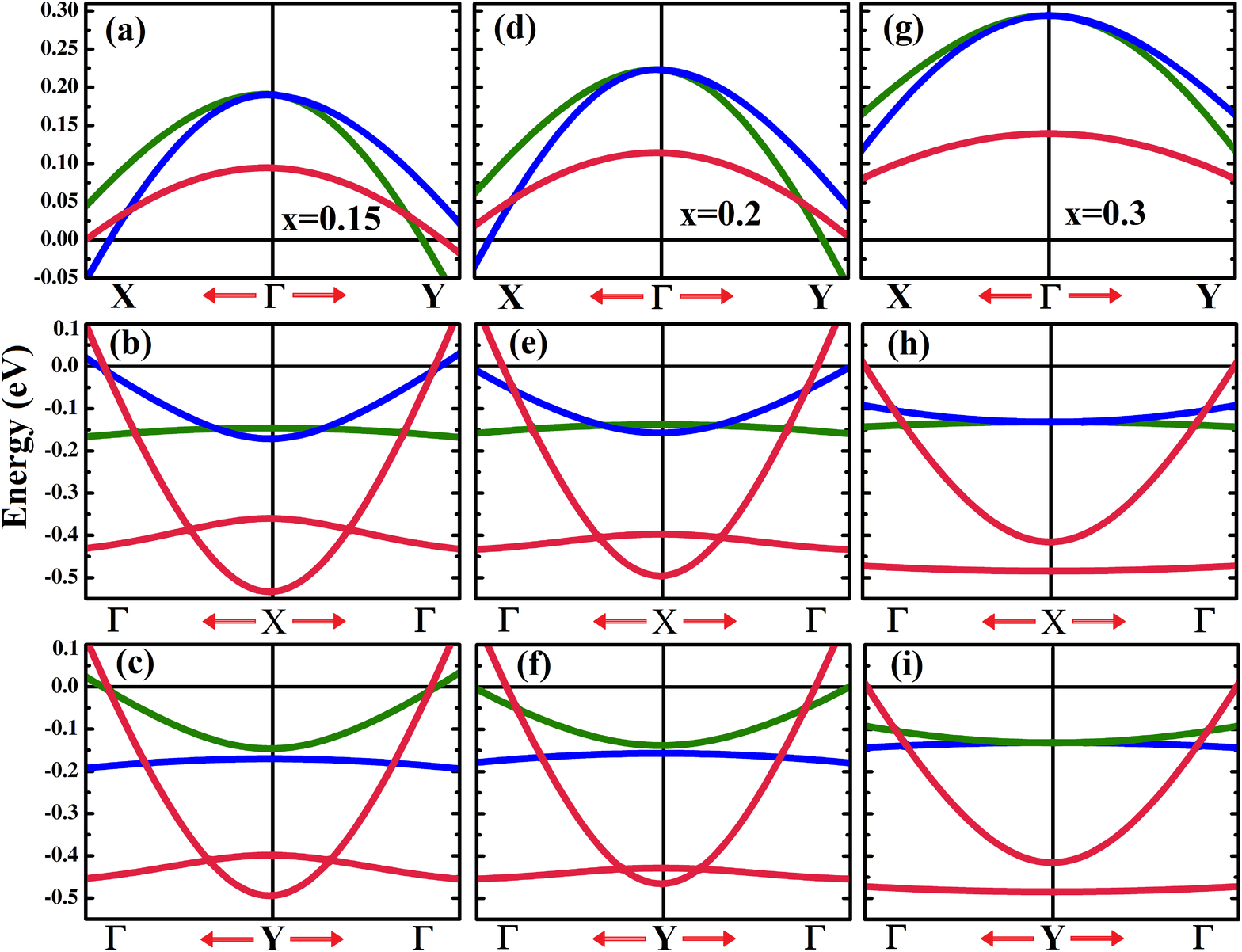}
  \caption{(Colour online) Calculated band structures of Ba$_{1-x}$Na$_x$Fe$_2$As$_2$ system within VCA method 
      in the low temperature orhorhombic phase 
      around $\Gamma$ (1st row), X (2nd row) and Y (3rd row) points for 
      (a, b, c) $x=0.15$ (d, e, f) $x=0.2$ and (g, h, i) $x=0.3$ respectively. Fermi level is denoted by 
      horizontal black line at zero energy. Orbital orders around all these high symmetry points are calculated.}
  \label{NaBS}
  \end{figure}

 First we calculate the band structures of various doped BaFe$_2$As$_2$ (Ba122) compounds as a function of doping concentrations using experimental lattice parameters ($a$, $b$, $c$ and $z_{As}$) in the orthorhombic as well as tetragonal phases. 
We present the band structures around high symmetry k points ($\Gamma$, X and Y) of K and Na doped (\textit {i.e}, hole doped) 
Ba122 systems in fig.\ref{KBS} and fig.\ref{NaBS} respectively for various doping concentrations as indicated in figures.
Different colours are used to designate various orbital projected bands near the Fermi level (FL). 
It is  an well established fact that the electronic structure near the FL of these Fe-based SCs are mainly dominated by Fe-d orbitals. It is quite evident from the two figures (fig.\ref{KBS} and fig.\ref{NaBS}) that as a consequence of hole doping, hole like bands around $\Gamma$ points (mainly d$_{xz}$ and d$_{yz}$ bands near FL) expands as well as move away from the FL with increasing hole doping 
   \begin{figure}[ht]
    \centering
    \includegraphics [height=8cm,width=8.5cm]{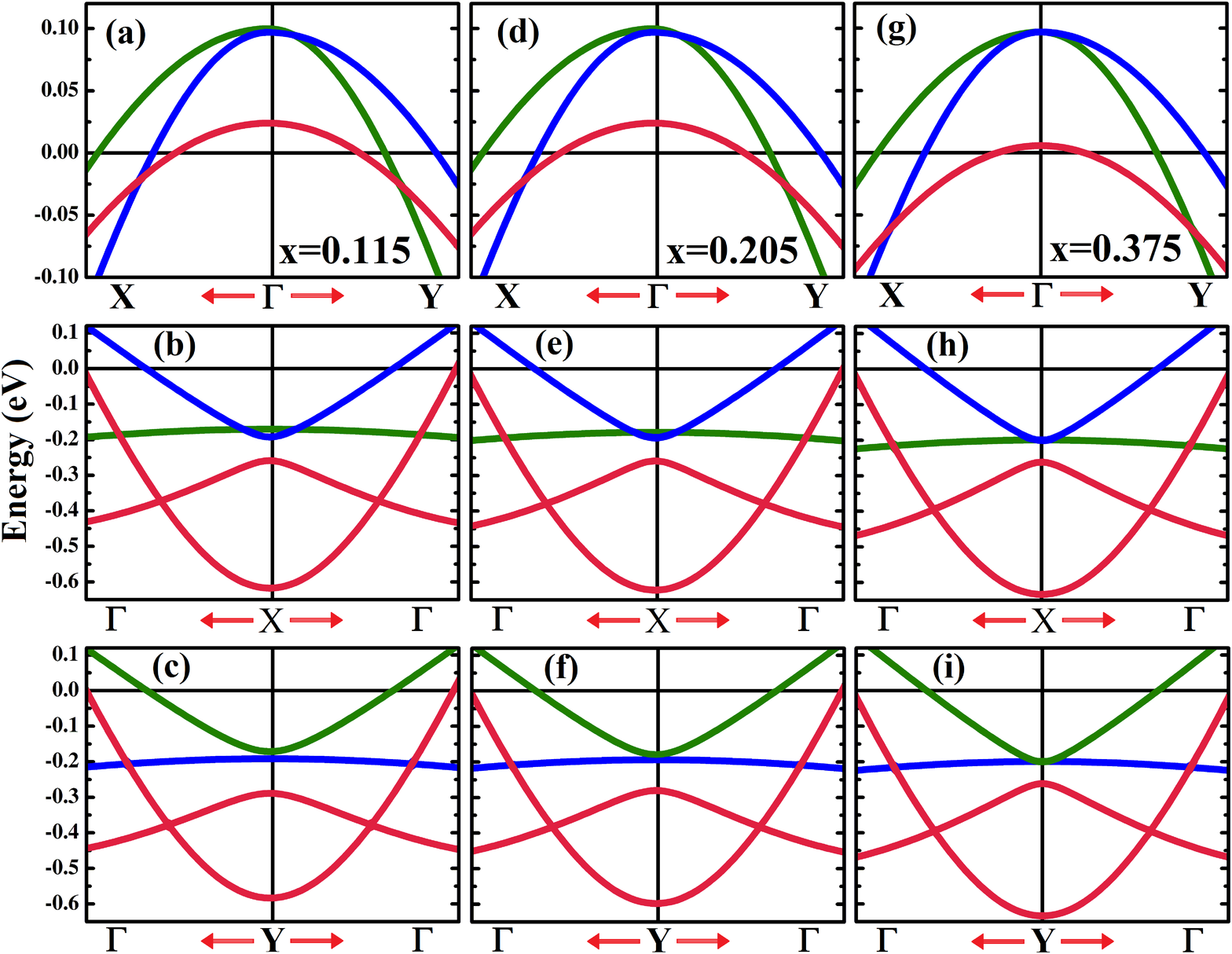}
    \caption{(Colour online) Calculated band structures of BaFe$_2$(As$_{1-x}$P$_x$)$_2$ system within VCA method 
    in the low temperature orthorhombic phase 
    around $\Gamma$ (1st row), X (2nd row) and Y (3rd row) points for 
    (a, b, c) $x=0.115$ (d, e, f) $x=0.205$ and (g, h, i) $x=0.375$ respectively. Fermi level is denoted by 
    horizontal black line at zero energy. Orbital orders around all these high symmetry points are calculated.}
    \label{PBS}
    \end{figure}
concentration. On the other hand, electron like bands (mainly d$_{xy}$ bands near FL) become flat around X, Y points as well as moves towards the FL. 
This whole scenario is consistent with the hole doping picture where hole like Fermi surfaces expand and electron like Fermi surfaces shrink with increasing doping concentration. Figs \ref{NaBS}, \ref{KBS} are ideal picture of inter band nesting whereas clear degradation of intra band nesting with doping. We also depict the calculated band structures around high symmetry 
points ($\Gamma$, X and Y) of P doped (iso-electronic) Ba122 system in the orthorhombic and tetragonal phases in Fig.\ref{PBS} 
for various doping concentrations. Fig.\ref{PBS} reveals that modifications in the position of the bands 
near FL due to P doping at As sites in Ba122 system is comparatively weaker than hole doped 122 materials. However there is some visible moderation in the d$_{xy}$ band near FL 
around $\Gamma$ point with the variation of P doping concentration \--- exactly opposite behaviour with doping compared to the hole doped materials show in the top rows of figures 1, 2. This indicates effectively electron doping.  
Iso-electronic P doping at As site is like in-plane substitution (substitution in the Fe-As layer) contrasting with 
K or Na substitution at Ba site which is out of plane substitution (in between two Fe-As layer). 
Therefore, P doping at As site affects the d$_{xy}$ bands more drastically than the other bands (mainly d$_{yz}$ and d$_{xz}$ bands) near FL. As a result d$_{xy}$ band (being planer in nature) gets modified remarkably with increasing doping concentration. As a side remark we would like to mention that in Fig. \ref{PBS} (g) corresponding to $x$= 0.375 of P doping at the As site leads to a Lifshitz transition as the planer d$_{xy}$ band about to cross the FL. 
Clearly, with doping the d$_{xy}$ band intersects the FL (the black solid line at zero) at two closer and closer k-points, leading to the difference between the two k points $\delta k \to 0$, which corresponds to vanishing of the inner FS, carrying the vital signature of Lifshitz transition.  This subject is not in the purview of the
present paper and interested readers may follow a detailed manifestations in ref. \cite{LT-hng}.
However, in case of K and Na doping at Ba site (hole doping) 
which is by nature an out of plane substitution, the planer d$_{xy}$ band near FL is comparatively less affected 
but there are notable moderations in the d$_{yz}$ as well as d$_{xz}$ bands near FL due to hole doping.
From figs.\ref{KBS}, \ref{NaBS}, \ref{PBS}, we calculate the band 
energies of d$_{xz}$ and d$_{yz}$ bands around all high symmetry k points: X, Y, $\Gamma$ for K, Na and P 
doped Ba122 systems. 
In fig.\ref{FeFedop}, sum of the band energies \textit{ i.e.}, E$_{d_{xz}}$ and E$_{d_{yz}}$ around  X, Y, $\Gamma$ 
points as a function of doping concentration for various doped Ba122 systems are presented. It has been explicitly 
shown that these band energies (E$_{d_{xz}}$ and E$_{d_{yz}}$) follow the same behaviour as that of the two 
Fe-Fe bond lengths (in the orthorhombic phase of these Ba122 systems, there are two different Fe-Fe bond lengths indicating stripe SDW order) as 
a function of doping concentration for K, Na as well as P doped Ba122 systems.
  \begin{figure}
   \centering
   \includegraphics [height=8cm,width=8cm]{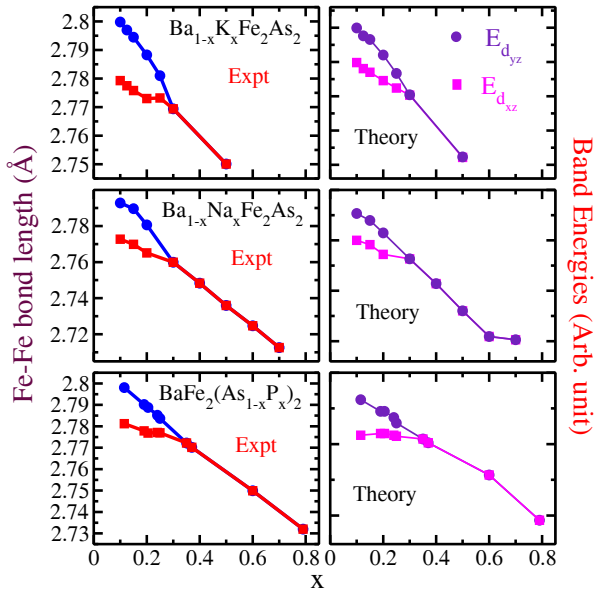}
   \caption{(Colour online) Experimental two Fe-Fe bond lengths (1st column) and sum of the calculated band energies of $d_{yz}$ and $d_{xz}$ bands around high symmetry points (2nd column) as a function of doping for K (1st row), 
            Na (2nd row) and P (3rd row) doped BaFe$_2$As$_2$ systems.}
   \label{FeFedop}
   \end{figure}
 \begin{figure}
  \centering
  \includegraphics [height=8cm,width=8.5cm]{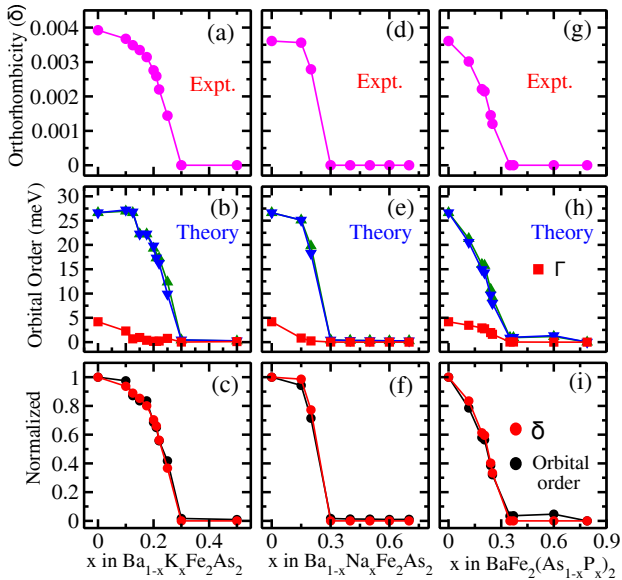}
  \caption{(Colour online) Experimental orthorhombicity parameter of different 122 materials (1st row), 
   theoretically calculated orbital orders 
     around $\Gamma$, X, Y points (2nd row) and normalized orthorhombicity parameter as well as orbital order (3rd row) as a function of doping for (a, b, c) K doped (d, e, f) Na doped (g, h, i) P doped BaFe$_2$As$_2$ system respectively.}
  \label{OOdop}
  \end{figure}
In fig.\ref{FeFedop} (1st column), we also display the experimental values of Fe-Fe bond lengths as a function of doping concentration 
for various doped Ba122 systems taken from references \cite{Avci,Allred,Avci2}. We also estimate the differences 
of those band energies around $\Gamma$, X, Y points 
which is defined as orbital order (orbital order = $\sum_i$ E$^i_{d_{xz}}$-E$^i_{d_{yz}}$, where $i=\Gamma, X, Y$) 
for a number of 122 systems for each doping concentration (see fig.\ref{OOdop}b, \ref{OOdop}e, \ref{OOdop}h). In fig.\ref{OOdop},
we show the experimentally measured variation of orthorhombicity parameter [defined as $\delta=(a-b)/(a+b)$)] 
of Ba122 systems with K, Na as well as P doping concentration (1st row). 
Estimated orbital order as a function of doping concentration around X, Y and $\Gamma$ 
points for the same three systems are also presented in fig. \ref{OOdop}b, \ref{OOdop}e, \ref{OOdop}h. 
We also display the normalized orthorhombicity parameter 
as well as orbital order for the same three systems as a function of doping concentration 
in fig. \ref{OOdop}c, \ref{OOdop}f, \ref{OOdop}i to visualize 
the universal mapping of structural transition to the orbital ordering. 
One can easily see from those figures (fig. \ref{OOdop}c, \ref{OOdop}f, \ref{OOdop}i) that 
orbital order and orthorhombicity parameter quantitatively follow identical behaviour with doping concentration 
for all the studied 122 systems (K, Na and P doped Ba122). Therefore, through Figures \ref{FeFedop}, \ref{OOdop} we show that the structural parameters like two inequivalent Fe-Fe distances of the low temperature orthorhombic phase is mapped to the electron band energy and the orthorhombicity parameter which characterizes structural transition of a compound is mapped to `orbital order' for three different types of doped 122 materials hole, electron and iso-electronic ones. The same mapping will now be demonstrated as far as temperature variations are concerned through figures \ref{undopedBST},\ref{2BST},\ref{Fe-Fe},\ref{OOT}.
  \begin{figure}
   \centering
   \includegraphics [height=8cm,width=8.5cm]{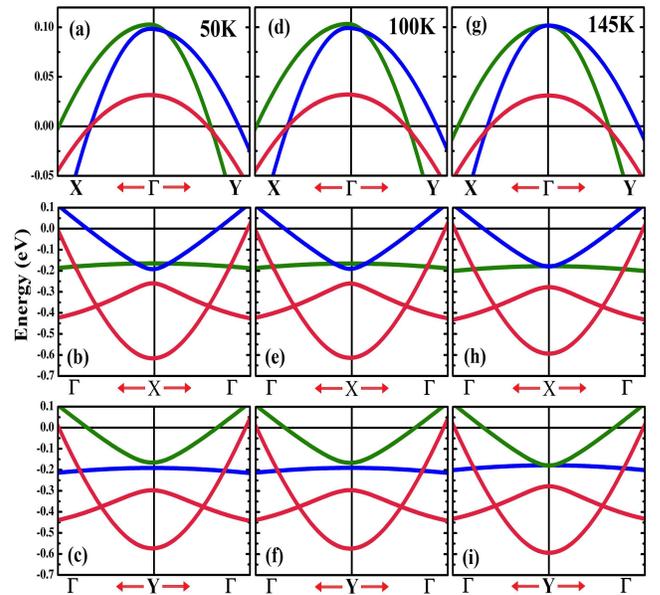}
   \caption{(Colour online) Calculated band structures of undoped BaFe$_2$As$_2$
    around $\Gamma$ (1st row), X (2nd row) and Y (3rd row) points for 
       (a, b, c) T = 50 K (d, e, f) T = 100 K and (g, h, i) T = 145 K. Fermi level is denoted by the
       horizontal black line at zero energy. Orbital orders around all these high symmetry points are calculated.}
   \label{undopedBST}
   \end{figure}
     \begin{figure}
      \centering
      \includegraphics [height=8cm,width=8.5cm]{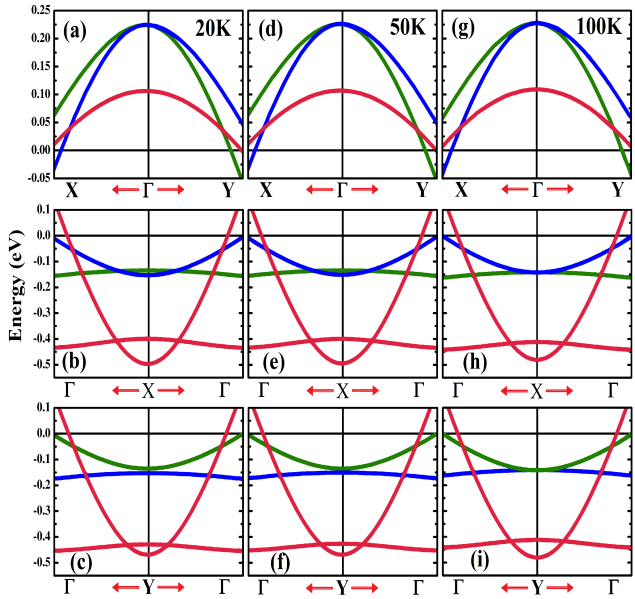}
      \caption{(Colour online) Calculated band structures of Ba$_{0.8}$K$_{0.2}$Fe$_2$As$_2$ within VCA   
          around $\Gamma$ (1st row), X (2nd row) and Y (3rd row) points for 
             (a, b, c) at T = 20 K, (d, e, f) at T = 50 K and (g, h, i) at T = 100 K respectively. 
             Fermi level is denoted by 
         the horizontal black line at zero energy. Orbital orders around all these high symmetry points are calculated, see text.}
      \label{2BST}
      \end{figure}
        \begin{figure}
         \centering
         \includegraphics [height=8cm,width=8cm]{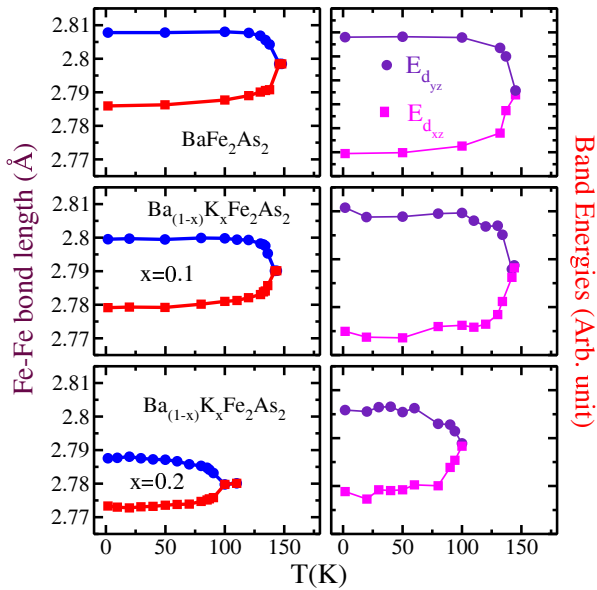}
         \caption{(Colour online) Experimental Fe-Fe bond lengths (1st column) and sum of the calculated band energies of $d_{yz}$ and $d_{xz}$ bands around high symmetry points (2nd column) as a function of temperature for undoped BaFe$_2$As$_2$ (1st row), 
         10\% (2nd row) and 20\% (3rd row) K doped BaFe$_2$As$_2$ respectively.}
         \label{Fe-Fe}
         \end{figure}
\par We further extend our studies in case of temperature dependent structural transitions for 
undoped as well as K-doped Ba122 systems to prove the universal mapping of orbital order to orthorhombicity parameter as far as there thermal variations are concerned. 
We calculate the band structures of undoped as well as K-doped Ba122 systems (10\% and 20\% K doping) 
using temperature and doping dependent experimentally determined 
structural parameters [$a(x,T), b(x,T), c(x,T)$ and z$_{As}(x,T)$]. Calculated band structures around X, Y, $\Gamma$ 
points at various temperatures for undoped 
and 20\% K doped Ba122 systems are presented in fig.\ref{undopedBST} and fig.\ref{2BST} respectively. We follow the same procedure of calculating temperature dependent two inequivalent Fe-Fe distances and orbital order as earlier in case of doping dependent studies above.
Unlike doping dependent band structures, temperature dependent band structures show weak modification with 
temperature near the FL. But a more closer look to the band structures for various temperatures, reveals some 
important informations about structural transition and its link to electronic structures (see for example, visible modifications at around X and Y points in figures \ref{undopedBST}, \ref{2BST} (b\-- i)). 
Here also we calculate the sum of band energies E$_{d_{xz}}$ and E$_{d_{yz}}$ around X, Y, $\Gamma$ points at each temperature and compare these values with temperature dependent experimentally measured Fe-Fe bond lengths. In fig.\ref{Fe-Fe}, experimentally measured temperature dependent Fe-Fe bond lengths taken from ref \cite{Avci} are presented in column one for undoped as well as 10\% and 20\% K doped Ba122 systems. In the second column of fig. \ref{Fe-Fe}, temperature variation of band energies (E$_{d_{xz}}$ and E$_{d_{yz}}$) are presented for the same three systems. Temperature dependence of band energies follow the same experimentally observed thermal behaviour 
of two Fe-Fe distances. This should also be contrasted with Fig.\ref{FeFedop} where doping dependent 
Fe-Fe distances are also followed by the band energies. This is an example of universal 
correlation of experimental structural parameters with electronic structure.
We also calculate the orbital orders around X, Y, $\Gamma$ points as a function of temperature 
for these systems and present in fig.\ref{OOT}b, \ref{OOT}e and \ref{OOT}h respectively.
  \begin{figure}
   \centering
   \includegraphics [height=8cm,width=8.5cm]{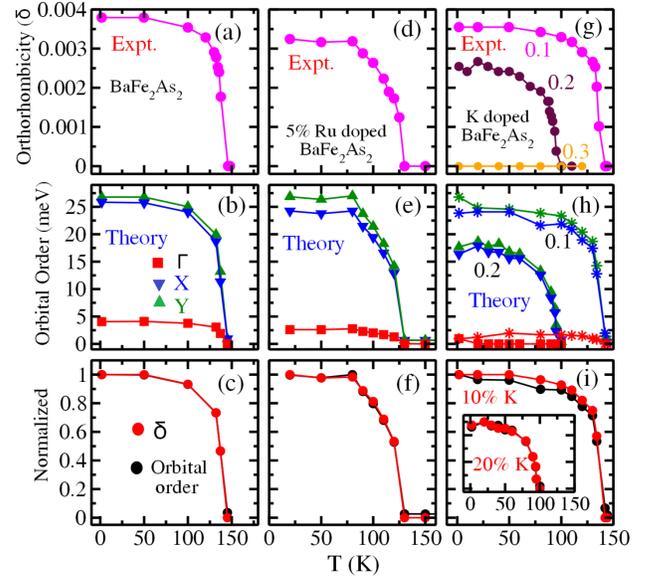}
   \caption{(Colour online) (a, b, c) (Column wise) Experimental orthorhombicity parameter, calculated orbital orders 
   around high symmetry $\Gamma$, X, Y points and normalized orthorhombicity parameter as well as orbital order (as defined in the text) as a function of temperature for undoped BaFe$_2$As$_2$ respectively. (d, e, f) Experimental orthorhombicity parameter, calculated orbital orders around $\Gamma$, X, Y points and normalized orthorhombicity parameter as well as orbital order (as defined in the text) as a function of temperature for 5\% Ru doped BaFe$_2$As$_2$ system respectively. (g, h, i) Experimental orthorhombicity parameter (including 30\% K doped BaFe$_2$As$_2$ in the top figure (g)), calculated orbital orders around $\Gamma$, X, Y points and normalized orthorhombicity parameter as well as orbital order (as defined in the text) as a function of temperature for 10\%, 20\% K doped BaFe$_2$As$_2$ respectively.}
   \label{OOT}
   \end{figure}
  \begin{figure}[ht]
   \centering
   \includegraphics [height=6cm,width=8cm]{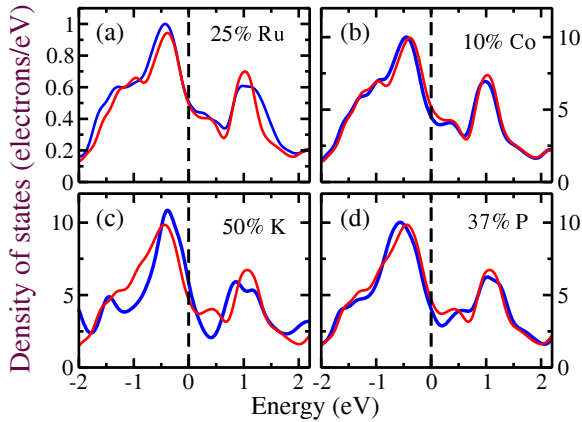}
   \caption{(Colour online) Calculated density of states of (a) 25 \% Ru doped (iso-electronic doping) (b) 10\% Co doped (electron doping), (c) 
   50\% K doped (hole doping) and (d) 37\% P doped (iso-electronic doping) BaFe$_2$As$_2$ respectively. To compare the 
   shift of chemical potential $\mu$ due to doping, density of states of the undoped 122 is also depicted
   with a solid red line. Vertical dashed line at 0 eV indicate the Fermi level.}
   \label{DOSa}
   \end{figure}
Experimentally observed orthorhombicity parameter as a function of temperature 
is also depicted in fig.\ref{OOT}a, \ref{OOT}d, \ref{OOT}g for these 
systems. We also show in fig.\ref{OOT}c, \ref{OOT}f, \ref{OOT}i, that normalized 
orbital order maps with the normalized orthorhombicity parameter as a function of temperature.
Thus, it is very clear that the structural transition in 122 Fe-based SC is due to the 
orbital ordering between d$_{xz}$, d$_{yz}$ bands and thus electronic in origin. 
Our studies so far also suggest that the electronic structure is highly sensitive to structural parameters and there are universal correlations between structural parameters to the electronic structure in these classes of superconductors as evident from 
the fact that two band energies (E$_{d_{xz}}$ and E$_{d_{yz}}$) follow the same temperature as well as doping dependence as that of the two Fe-Fe bond lengths.

\par
\begin{figure}[h!]
   \centering
   \includegraphics [height=9cm,width=8.5cm]{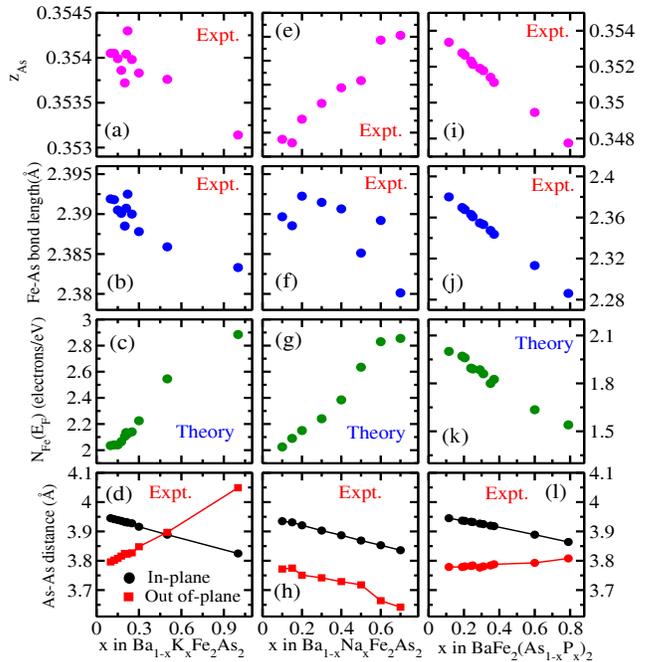}
   \caption{(Colour online) Experimentally determined z$_{As}$ (1st row), Fe-As bond lengths (2nd row), calculated DOS of Fe atoms at the Fermi level (3rd row) and experimental in-plane as well as out of plane As-As distances (4th row) as a function of doping concentration for (a-d) K (e-h) Na and (i-l) P doped BaFe$_2$As$_2$ respectively.}
   \label{DOS}
   \end{figure}

\begin{figure}[h!]
   \centering
   \includegraphics [height=9cm,width=7.5cm]{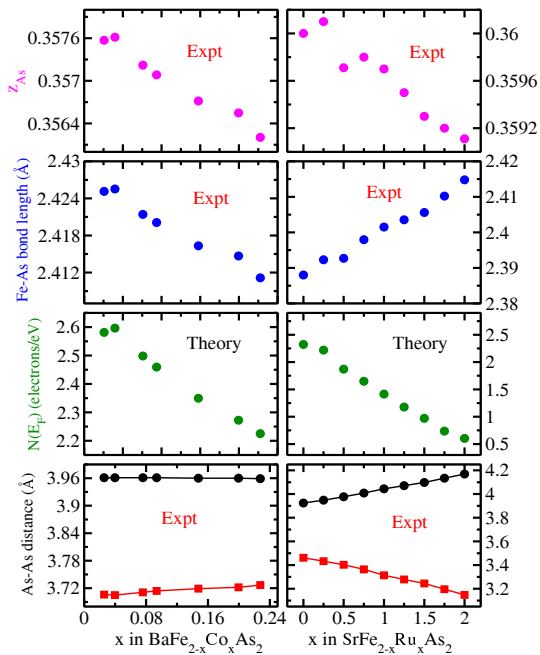}
   \caption{(Colour online) Experimentally determined z$_{As}$ (1st row), Fe-As bond length (2nd row), theoretically calculated DOS of Fe atoms at the Fermi level (3rd row) and experimental in-plane as well as out of plane As-As distances (4th row) as a function of doping concentration for Co (1st column) and Ru (2nd column) doped BaFe$_2$As$_2$ respectively.}
   \label{CoRuDOS}
\end{figure}

 In order to find some universal correlation among other structural parameters and electronic structures, we further investigate various doped Ba122 systems in the orthorhombic and tetragonal phases. First, we calculate the density of states for different doping concentrations for undoped and various doped 
Ba122 systems like Ba$_{1-x}$K$_x$Fe$_2$As$_2$, Ba$_{1-x}$Na$_x$Fe$_2$As$_2$, BaFe$_{2-x}$Co$_x$As$_2$, 
BaFe$_{2-x}$Ru$_x$As$_2$ and BaFe$_2$(As$_{1-x}$P$_x$)$_2$. In fig.\ref{DOSa}, we depict the calculated density of states of undoped, 10\% Co doped, 50\% K doped and 32\% P doped Ba122 systems near the FL. 
In fig.\ref{DOSa}b, \ref{DOSa}c, \ref{DOSa}d, we also exhibit the density of states of the undoped system
with the doped one for comparison. We precisely choose to present the density of states of the doped Ba122 
compounds with optimal doping concentrations at which these materials have the highest T$_c$.
It is evident from fig.\ref{DOSa}, that the chemical potential (Fermi level) shifts in the opposite directions in case of hole and electron doping as expected. From this figure it also turns out that the case of iso-electronic P doping is very similar to the case of electron doping. Up to 25$ \%$ doping although it is not very clear whether Ru doping correspond to hole or electronic doping, at higher doping (not shown here) it behaves like hole doping. Main purpose of this figure is to provide guidelines to the readers regarding the possible behaviour of the DOS presented in the third rows of the figures\ref{DOS} and \ref{CoRuDOS}. 
In fig.\ref{DOS} and fig.\ref{CoRuDOS}, we display 
the experimentally measured various structural parameters (taken from references 
\cite{Avci,Allred,Avci2,Acta,Sefat,Ni}) and theoretically 
computed electronic structures as a function of doping for a number of doped Ba122 materials in the 
orthorhombic and tetragonal phases. In the first 
and second rows of fig.\ref{DOS}, we exhibit the experimental variation of z$_{As}$ as well as Fe-As bond 
lengths with doping concentration respectively for K, Na, P doped Ba122 materials. 
In fig.\ref{DOS} c, g, k (3rd row), calculated density of states at the FL (N(E$_F$)) 
for these compounds are presented as a function of doping concentration.

In the last row, we exhibit the experimentally estimated in-plane and out of plane As-As distances as a function of
doping concentration for these three systems. Although in all the above three systems, doping is not at the Fe site,
there are significant variation of structural parameters with doping concentration. However, we also consider the case of active site doping \textit{i.e.,} Co and Ru doping at the Fe site. In fig.\ref{CoRuDOS}, we present the variation of experimental z$_{As}$ as well as experimental Fe-As bond length with doping concentration for Co, Ru doped Ba122 in 1st and 2nd rows respectively. Calculated density of states at the FL (N(E$_F$)) for these two systems as a function of doping concentration are exhibited in the 3rd row of fig.\ref{CoRuDOS}. In the last row of
fig.\ref{CoRuDOS}, we depict the variation of in plane as well as out of plane As-As distances with doping concentration for Co and Ru doped Ba122 systems.
From fig.\ref{DOS} and fig.\ref{CoRuDOS}, we observe that, there are diversities and universalities
in the behaviour of structural parameters as well as electronic structure with doping concentration.
In case of hole doping (K and Na doped Ba122 systems),
density of states at the FL inversely follow the variation of Fe-As bond length with doping. On the contrary,
density of states at the FL follow the same behaviour as that of the Fe-As bond length with doping for the case of electron doped (Co doped) Ba122 system. Moreover, the case of iso-electronic P doping in As site is very similar to the case of electron doping as evident from the doping dependent behaviour of experimental z$_{As}$, Fe-As bond length as well as calculated density of states at the FL. Although the case of iso-electronic Ru doping at Fe site is similar to the hole doping scenario where density of states at the FL follow the same doping variation as that of the Fe-As bond length; there is in general an universal link between the variation of density of states 
at the FL with that of the experimental z$_{As}$ with doping
and it turns out that z$_{As}$ is inversely (directly) proportional to Fe-As bond distances
(with exceptions of K doped BaFe$_2$As$_2$) for hole (electron) doped materials as a function of doping.

\par Now we shed more light into the universality and diversities in the behaviour of electronic structure and structural parameters of 122 Fe based compounds. In plane and out of plane As-As distances also play very important and crucial role in controlling the electronic structures. First, we study the case of out of plane substitution \textit{i.e.,} K and Na doping at Ba site. Size of the Ba atom (2.22\AA) [atomic radius in metalic bonding] is larger than Na atom (1.86\AA) but smaller than K atom (2.27\AA). Because of this reason, with the substitution 
of K atom in place of Ba atom, out of plane As\--As distance increases and exactly opposite behaviour in the out of plane As-As distance is observed in the case of Na substitution in Ba site. It should also be noted that the c-axis 
increases with the increasing doping concentration for both Na and K doping (in case of Na doped Ba122 materials 
c-axis increases with doping up to certain doping concentration and after that it decreases). This qualitatively explains the observed behaviour of z$_{As}$ with doping for K and Na doped Ba122 systems. 
On the other hand, in both the hole doped cases, in plane As-As distance decreases with increasing doping concentration following the doping variation of Fe-As bond length. Next we investigate the case of in plane substitutions (Co/Ru substitution at Fe site and P substitution at As site).
Size of Fe atom (1.26\AA) is smaller than Ru atom (1.34\AA) but larger than Co atom (1.25\AA). Therefore, Fe-As 
bond length decreases with Co doping but increases with doping in case of Ru substitution in place of Fe. 
This is consistent with the behaviour of in-plane as well as out of plane As-As distances as a function of doping. 
Since the size of P atom (1\AA) is smaller than As atom (1.15\AA), Fe-As bond length decreases with increasing doping 
concentration just like the case of Co doped (electron doped) system. These are also consistent with the doping dependent 
variation of out of plane As-As distance (in all the cases of in plane substitution c-axis decreases with increasing 
doping concentration). Moreover, in plane As-As distance in all these cases (in plane substitution) follow 
the same doping dependencies as that of the Fe-As bond length. 
Thus all these structural parameters are crucially controlling the electronic structures near the FL of the 122 compounds and  follow universal characteristics features. A shuttle dependence of chemical size, atomic number of the host/substituent, in-plane or out of plane nature of substitution determines ``aggregate electronic motion''
which is thus detrimental of electronic structure, band structure and FS. This governs various electronic properties
of 122 Fe-based materials like topological transition, magneto-structural transition, nematicity and 
possibly superconductivity \cite{Sen,LT-hng,Ren}  
\section{Conclusion}

A detailed first principles electronic structure calculation on a number of doped 122 family of 
Fe-based SCs are presented. Experimentally Rietveld refined structural parameters like lattice parameters ($a, b, c$) 
together with z$_{As}$ as a function of doping as well as temperature are used as fixed inputs in our calculations. 
Various kinds of dopings like; electron doping, hole doping, iso-electronic doping are considered 
(within VCA) in this study. Electronic band structure which essentially represent an aggregate electronic motion (band motion) carries the ``finger prints'' of any structural or magnetic modifications with temperature or doping. This physical aspect is used to calculate orbital ordering and various correlations of intrinsic structural parameters to electronic structure (e.g, two Fe-Fe distances maps with E$_{d_{xz}}$, E$_{d_{yz}}$ band energies, Fe-As or z$_{As}$, As-As distances etc. to N(E$_F$)). Shuttle interplay of structural parameters with electronic structure is employed to explain as to why iso-electronic doping P in place of As and Ru in place of Fe corresponds
to electron and hole dopings respectively.  It has been shown very rigorously and quantitatively for a number of Fe-based SCs that the orbital ordering between the d$_{xz}$, d$_{yz}$ orbitals is the origin of structural transition. 
The calculated orbital ordering from electronic structure calculations when normalized its variation with temperature 
as well as doping identically follows (quantitatively) with that of the experimental orthorhombicity parameter. 
This conclusively proves that the structural transition in 122 family of Fe-based SCs is not lattice driven but 
electronic one. Therefore, our work supports the electronic origin of the nematic phase observed in 122 family 
of Fe-based materials. An universal correlation among the structural parameters and the electronic 
structure is described in this work.  

\section{Acknowledgements} One of us H. Ghosh, thanks D. Vanderbilt for useful discussions on VCA and its applicability. He also gratefully acknowledges fruitful discussions with R. Osborn and J. M. Allred on structural parameters and electronic structure. Detailed discussions on application of experimental temperature and doping dependent structural parameters to DFT with Harald O. Jeschke, Aron Walsh are very gratefully acknowledged by H. Ghosh. We thank Dr. P. A. Naik, Dr. P. K. Gupta and Dr. P. D. Gupta for their encouragements in this work. Some of us (S. Sen, A. Ghosh) acknowledges the HBNI, RRCAT for financial support and encouragements. 


\begin{thebibliography}{9}
 
\bibitem{Kazuhiko} Kazuhiko Kuroki, Seiichiro Onari, Ryotaro Arita, Hidetomo Usui, Yukio Tanaka, Hiroshi Kontani, and Hideo Aoki. Phys. Rev. Lett. {\bf 101}, 087004 (2008).
\bibitem{Mazin} I. I. Mazin, D. J. Singh, M. D. Johannes and M. H. Du. Phys. Rev. Lett. {\bf 101}, 057003 (2008).
\bibitem{Hirschfeld} P. J. Hirschfeld, M. M. Korshunov, and I. I. Mazin. Rep. Prog. Phys. {\bf 74}, 124508 (2011).
\bibitem{Reid} J. P. Reid, M. A. Tanatar, A. Juneau-Fecteau, R. T. Gordon, S. Rene de Cotret, N. Doiron-Leyraud, T. Saito, H. Fukuzawa, Y. Kohori, K. Kihou, C. H. Lee, H. Iyo, H. Eisaki, R. Prozorov, and L. Taillefer. Phys. Rev. Lett. {\bf 109}, 087001 (2012).
\bibitem{Chu} J.-H. Chu, J. G. Analytis, K. De Greve, P. L. McMahon, Z. Islam, Y. Yamamoto, and I. R. Fisher. Science {\bf 329}, 824 (2010).
\bibitem{Prost} J. Prost and P. G. de Gennes. The Physics of Liquid Crystals. Oxford Science Publicationq, 1974.
\bibitem{Fradkin} E. Fradkin, S. A. Kivelson, M. J. Lawler, J. P. Eisenstein, and A. P. Mackenzie. Annual Review of Condensed Matter Physics {\bf 1}, 153 (2010).
\bibitem{Song} Yu Song, Xingye Lu, D. L. Abernathy, David W. Tam, J. L. Niedziela, Wei Tian, Huiqian Luo, Qimiao Si, and Pengcheng Dai. 
Phys. Rev. B {\bf 92}, 180504(R) (2015)
\bibitem{Ni} N. Ni, M. E. Tillman, J.-Q. Yan, A. Kracher, S. T. Hannahs, S. L. Bud'ko, and P. C. Caneld. Phys. Rev. B  {\bf 78}, 214515 (2008).
\bibitem{Chu2} J.-H. Chu, J. Analytis, C. Kucharczyk, and I. R. Fisher. Phys. Rev. B {\bf 79}, 014506 (2009).
\bibitem{Luetkens} H. Luetkens, H-H Klauss, M. Kraken, F. J. Litterst, T. Dellmann, R. Klingeler, C. Hess, R. Khasanov, A. Amato, C. Baines, M. Kosmala, O. J. Schumann, M. Braden, J. Hamann-Borrero, N. Leps, A. Kondrat, G. Behr, J. Werner, and B Bluchner, Nature Mater. {\bf 8}, 305 (2009).
\bibitem{Parker} D. R. Parker, M. J. P. Smith, T. Lancaster, A. J. Steele, I. Franke, P. J. Baker, F. L. Pratt, M. J. Pitcher, S. J. Blundell, and S. J. Clarke. Phys. Rev. Lett. {\bf 104},0571007 (2010).
\bibitem{Margadonna} S. Margadonna, Y. Takabayashi, M. T. McDonald, K. Kasperkiewicz, Y. Mizuguchi, Y. Takano, A. N. Fitch, E. Suard, and K. Prassides. Chem. Comm. {\bf 43}, 5607 (2008).
\bibitem{McQueen} T. M. McQueen, A. J. Williams, P. W. Stephens, J. Tao, Y. Zhu, V. Ksenofontov, F. Casper, C. Felser, and R. J. Cava. Phys. Rev. Lett. {\bf 103}, 057002 (2009).
\bibitem{Hsu} Fong-Chi Hsu, Jiu-Yong Luo, Kuo-Wei Yeh, Ta-Kun Chen, Tzu-Wen Huang, Phillip M Wu, Yong-Chi Lee, Yi-Lin Huang, Yan-Yi Chu, Der-Chung Yan, and Maw-Kuen Wu. Proceedings of the National Academy of Sciences of the United States of America {\bf 105}, 14262 (2008).
\bibitem{Imai} T. Imai, K. Ahilan, F. L. Ning, T. M. McQueen, and R. J. Cava. Phys. Rev. Lett. {\bf 102}, 177005 (2009).
\bibitem{FernandesN}  R. M. Fernandes,	A. V. Chubukov,	and J. Schmalian. Nature Physics {\bf 10}, 97 (2014)
\bibitem{WZhang} Wenliang Zhang, J. T. Park, Xingye Lu, Yuan Wei, Xiaoyan Ma, Lijie Hao, Pengcheng Dai, Zi Yang Meng, Yi-feng Yang, Huiqian Luo, Shiliang Li. arXiv:1607.06549 (2016).
\bibitem{Chandra} P. Chandra, P. Coleman, and A. I. Larkin. Phys. Rev. Lett. {\bf 64}, 88 (1990).
\bibitem{Fang} C. Fang, H. Yao, W. F. Tsai, J. P. Hu, and S. A. Kivelson. Phys. Rev. B, {\bf 77},224509 (2008).
\bibitem{Xu} C. Xu, M. Muller, and S. Sachdev. Phys. Rev. B {\bf 78}, 020501 (2008).
\bibitem{Qi} Y. Qi and C. Xu. Phys. Rev. B {\bf 80}, 094402 (2009).
\bibitem{Fernandes} R. Fernandes, L. VanBebber, S. Bhattacharya, P. Chandra, V. Keppens, D. Mandrus, M. McGuire, B. Sales, A. Sefat, and J. Schmalian. Phys. Rev. Lett. {\bf 105},157003  (2010).
\bibitem{Paul} I. Paul. Phys. Rev. Lett. {\bf 107}, 047004 (2011).
\bibitem{Cano} A. Cano. Phys. Rev. B {\bf 84}, 012504 (2011).
\bibitem{Fernandes2} R. M. Fernandes, A. V Chubukov, J. Knolle, I. Eremin, and J. Schmalian, Phys. Rev. B {\bf 85}, 024534 (2012).
\bibitem{Fanfarillo} L. Fanfarillo, A. Cortijo, and B. Valenzuela. Phys. Rev. B  {\bf 91}, 214515 (2015).
\bibitem{Sen} Smritijit Sen and Haranath Ghosh, J. Alloys Comp. {\bf 675},  416 (2016).
\bibitem{Kruger} Frank Kruger, Sanjeev Kumar, Jan Zaanen, and Jeroen van den Brink. Phys. Rev. B {\bf 79}, 054504 (2009).
\bibitem{Lee} Chi-Cheng Lee, Wei-Guo Yin, and Wei Ku. Phys. Rev. Lett. {\bf 103}, 267001 (2009).
\bibitem{Chen} C.-C. Chen, J. Maciejko, A. Sorini, B. Moritz, R. Singh, and T. Devereaux. Physical Review B, {\bf 82},100504 (2010).
\bibitem{Lv} Weicheng Lv, Frank Kruger, and Philip Phillips, Phys. Rev. B, {\bf 82}, 045125 (2010).
\bibitem{Onari} S. Onari and H. Kontani. Phys. Rev. Lett. {\bf 109}, 137001 (2012).
\bibitem{Zhai} H. Zhai, F. Wang, and D.-H. Lee. Phys. Rev. B {\bf 80}, 064517 (2009).
\bibitem{CASTEP} S. J. Clark, M. D. Segall, C. J. Pickard, P. J. Hasnip, M. J. Probert, K. Refson, M. C. Payne, Zeitschrift fuer Kristallographie {\bf 220}(5-6) (2005) 567.
\bibitem{PBE} J. P. Perdew, K. Burke and M. Ernzerhof, Phys. Rev. Lett. {\bf 77} (1996) 3865.
\bibitem{Acta} Shilpam Sharma, A. Bharathi, K. Vinod, C. S. Sundar, V. Srihari, Smritijit
Sen, Haranath Ghosh, Anil K. Sinha and S. K. Deb, Acta. Cryst. B {\bf 71} (2015) 61.
\bibitem{sust} Smritijit Sen, Haranath Ghosh, A.K. Sinha, A. Bharathi, Supercond. Sci. Technol. {\bf 27} (2014) 122003.
\bibitem{zAs} Deepa Kasinathan, Alim Ormeci, Katrin Koch, Ulrich Burkhardt, 
Walter Schnelle, Andreas Leithe-Jasper and Helge Rosner, New J. Phys. {\bf 11}  025023 (2009).
\bibitem{DJSingh}D. J. Singh, Phys. Rev. B {\bf 78}  094511 (2008).
\bibitem{Zhang} L. Zhang and D. J. Singh, Phys. Rev. B {\bf 79}  174530 (2009).
\bibitem{Yin}Z.P. Yin, S. Leb\'egue, M.J. Han, B.P. Neal, S.Y. Savrasov, W.E. Pickett, Phys. Rev.
Lett. {\bf 101}  047001 (2008).
\bibitem{Sn} Haranath Ghosh and S. Sen, J. Alloys Comp. {\bf 677},  245 (2016).
\bibitem{Mazin2} I. Mazin, Phys. Rev. B {\bf 78},  085104 (2008).
\bibitem{pla} Smritijit Sen, Haranath Ghosh, Phys. Lett. A {\bf 379},  843 (2015).
\bibitem{Bellaiche} L. Bellaiche and D. Vanderbilt, Phys. Rev.B, {\bf 61},  7877 (2000).
\bibitem{VCA} Smritijit Sen, and Haranath Ghosh, Eur. Phys. J. B (2016) DOI:10.1140/epjb/e2016-70446-2.
\bibitem{Avci} S. Avci, O. Chmaissem, D. Y. Chung, S. Rosenkranz, E. A. Goremychkin, J. P. Castellan, 
I. S. Todorov, J. A. Schlueter, H. Claus, A. Daoud-Aladine, D. D. Khalyavin, 
M. G. Kanatzidis and R. Osborn, Phys. Rev. B  {\bf 85}, 184507 (2012).
\bibitem{Allred} J. M. Allred {\it et al.}, Phys. Rev. B {\bf 90}, 104513 (2014).
\bibitem{Avci2} S. Avci, {\it et al.}, Phys. Rev. B {\bf 88}, 094510 (2013).
\bibitem{spinAIP} Smritijit Sen, and Haranath Ghosh, AIP Conference Proceedings {\bf 1728}, 020180 (2016)
\bibitem{Sefat} A. S. Sefat, R. Jin, M. A. McGuire, B. C. Sales, D. J. Singh and D. Mandrus, Phys. Rev. Lett. {\bf 101}  117004 (2008).
\bibitem{LT-hng} Haranath Ghosh and Smritijit Sen, arXiv:1607.05931 (2016).
\bibitem{Ren} Xiao Ren, Lian Duan, Yuwen Hu, Jiarui Li, Rui Zhang, Huiqian Luo, Pengcheng Dai, and Yuan Li, Phys. Rev. Lett. {\bf 115}  197002 (2015).
\end{thebibliography}
\end{document}